# Mobile neutron monitor for latitude cosmic ray monitoring

Kobelev  P.G., Maurchev  E.A., Yanke  V.G.

*The Pushkov Institute of Terrestrial Magnetism, Ionosphere and Radiowave Propagation of the Russian Academy of Sciences (IZMIRAN), Moscow, RF*
yanke@izmiran.ru, maurchev1987@gmail.com, kobelev@izmiran.ru

**Abstract** Neutron monitors are a standard tool for high-precision continuous monitoring of galactic cosmic ray flux variations arising from variations in heliospheric conditions and solar activity for space weather applications. These measurements form the basis for solving the inverse problem of determining the cosmic ray anisotropy vector beyond the magnetosphere. To support such studies, periodic latitude measurements are necessary to determine the coupling functions of primary and secondary cosmic rays variations. **The aim** of this work is to develop and characterize a modernized standard neutron monitor based on a CHM-15 boron thermal neutron counter and a data acquisition system designed for marine expeditionary studies of cosmic ray variations. Modern nuclear physics **experimental methods** and the principles of microprocessor-based data acquisition systems were used to solve this problem. **The results** of test trials and of continuous monitoring showed that the characteristics of the upgraded and standard neutron monitor are similar, and the ease of use, compactness, and stability allow us to **conclude** that the mobile neutron detector can be used in expeditionary conditions with limited access for maintenance personnel.

**Keywords:** cosmic rays, neutron component, neutron monitor, data acquisition system, latitude measurements of cosmic rays.

## 1. Introduction

Despite their seventy-year history [Simpson, 1957, 2000; Hatton et al., 1964; Carmichael, 1968], neutron monitors remain an important tool for measuring the intensity of cosmic particles with energies >400 MeV. The energy range of neutron monitors is an extension of the upper energy limit measured by cosmic ray detectors in outer space. Due to their high count rate, neutron monitors are capable to measure even the weak anisotropy associated with galactic cosmic rays. To date, all the most reliable information about anisotropy has been obtained from measurements by ground-based detectors.

A standard neutron monitor includes a neutron multiplier in the form of lead rings, in which atmospheric secondary particles with energies ≥10 MeV have a high probability of interacting with the lead, generating evaporative neutrons with energies in the MeV range. They are slowed to thermal energies in a 25 mm-thick polyethylene moderator, which surrounds a boron $^{10}B$ or helium $^{3}He$ thermal neutron counter. A standard neutron monitor is also surrounded by a 75 mm-thick polyethylene reflector, which serves to confine the neutrons produced in the lead within the monitor body and minimize the detector's response to the flux of external thermal neutrons due to their albedo.

In the early 2000s, a small expeditionary neutron monitor was developed based on the LND25382 helium counter [Moraal H. et al., 2001; Kuwabara et al, 2006; Krüger et al., 2008, 2010, 2013; Maghrabi et al. 2012, 2020; Medina et al., 2013; Heber et al., 2014; Poluianov et al. 2015]. However, this detector provides only ~9% of the count rate of the single-counter 1NM64 neutron monitor, which is ~1 pps sec$^{-1}$ at sea level at high latitudes. Such a detector can only be used at mountain stations [Poluianov et al. 2015], for example, at the Concordia station in Antarctica, where the neutron flux is almost two orders of magnitude higher than at sea level.

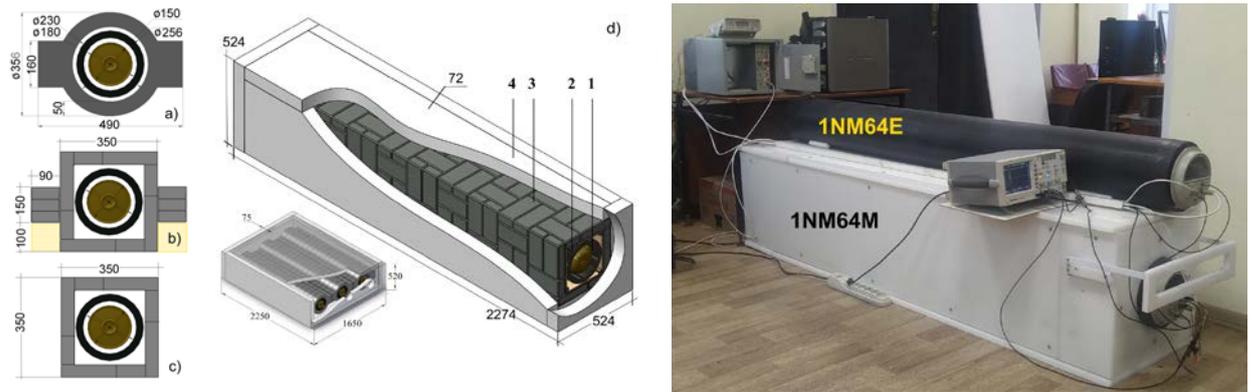

**Figure 1.** 1NM64M lead neutron monitor (schematic, 3D graphics in rectangular isometric view, and photo). The left panel shows cross-sections of the neutron detectors without the polyethylene reflector. The middle panel shows a 3D graphic of the detector (details in the text). The right panel shows a photo of the expedition neutron detector. The power supplies and data acquisition and processing system are located in the background.

A mobile neutron monitor based on the 3NM64 geometry is described in [Khamphakdee et al. 2025]. Lead-free neutron monitors [Nuntiyakul et al. 2018] are of particular interest for soil moisture studies [Zreda et al. 2008] and for nuclear safety monitoring [Kouzes et al. 2008]. However, using such detectors as a part of a global neutron detector network should be done with great caution, as their data are highly susceptible to environmental changes due to hydrogen-containing substances amount changes.

The global cosmic ray detector network is equipped with standard neutron monitors; for example, the Moscow station is equipped with a 24NM64 detector. This detector weighs 48 tons. Even placing the 6-counter expeditionary version (6NM64 in a 20-foot sea container) on ships presents challenges. When constructing new neutron detectors, the problem of manufacturing lead rings arises. We have developed and built a neutron detector based on SNM-15 large proportional counters, replacing the lead rings with standard lead blocks.

The aim of this work is to create a detector with the required characteristics for expeditionary measurements. The primary objective is to develop and describe an expeditionary neutron monitor, a data collection and processing system for expeditionary conditions with limited access for service personnel. In addition to secondary neutron intensity, the system must record necessary meteorological data and coordinates along the vessel's route.

## 2. Expedition detector configuration

The geometry of the standard NM64 neutron monitor was designed to maximize the effective count rate while minimizing detector weight. These detectors are built around large (ø150×1910 mm) SNM-15 or BP-28 proportional boron BF3 counters [https://www.lndinc.com]. The lead multiplier of the standard neutron monitor is formed from a set of specially manufactured lead rings (Figure 1, left panel, a), with a total weight of 1900 kg per counter and a count rate of ~430 ppm min-1 for the outermost counter in the section.

When creating the 1NM64M expeditionary neutron detector, it was necessary to replace the specially manufactured lead rings with standard lead blocks. Two options are possible.

The first option is shown in Figure 1 (left panel, b), where the dimensions and geometry of the lead ring of a standard monitor were closely replicated, right down to the arrangement of the "wings" which rest on a wooden supports. The total lead weight was 1914 kg, and the count rate was also ~430 ppm min-1.

The second option (under weight-limited conditions) is shown in Figure 1 (left panel, c); it is distinguished by the absence of "wings." The total lead weight is now 1320 kg, while the counting rate is ~290 ppm min-1, since the absence of "wings" reduces the effective detector area by ~⅓. Clarifying the characteristics requires creating a virtual model of the detector and conducting GEANT4 simulations. To organize the lead multiplier, 120 lead blocks measuring $200 \times 100 \times 50$ mm³ and 66 lead blocks measuring $180 \times 90 \times 50$ mm³ are required to form the "wings."

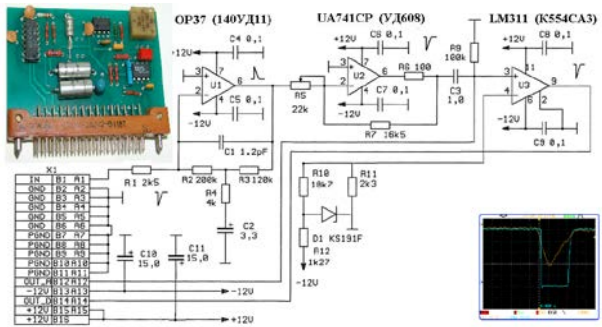

**Figure 2.** Discriminator amplifier with TTL output. Inserts – amplifier appearance and output analog and digital pulses.

Figure 1 (middle panel) shows an isometric projection of the detector model: 1 - thermal neutron counter, 2 - polyethylene cylindrical neutron moderator (24 mm), 3 - lead multiplier, 4 - polyethylene reflector (72 mm); The tab also shows the 3NM64 3-counter monitor. The right panel shows a photo of the completed expedition detector, above which is shown the 1NM64E epithermal neutron detector, consisting of: 1 – a thermal neutron counter and 2 – a polyethylene cylindrical neutron moderator.

### 3. Amplification tract

In counting mode, the amplifier does not have any special requirements other than the required output pulse amplitude and the time and temperature stability of its parameters. Therefore, the general requirements for the amplifier can be formulated as follows: simplicity, reliability, and high stability with respect to all destabilizing parameters (temperature, supply voltage).

The broadband amplifier (Figure 2) consists of two general-purpose operational amplifiers (OA). The first stage is a current-to-voltage converter, while the second serves for additional amplification.

The first stage is implemented using an inverting OA circuit with voltage feedback [Shilo, 1979]. The first stage's feedback circuit utilizes a technique whereby a large effective feedback resistance is created using resistors with relatively low resistance values. In this case, the OA output voltage is divided by a factor of 60 using a divider $R_3 - R_4$ and the presented feedback circuit works as a single resistor with a resistance $R_{oc} = R_2 (R_3/R_4) = 6$ MOhm. Let's perform an approximate analysis and calculation of the amplifier's first stage. When the OA is connected invertingly, the input current serves as the control signal, and a voltage is generated at its output. The main transfer function of this circuit is the transfer impedance. $R_{TR} = -R_{OA} = -R_2 (R_3/R_4) = -6$ MOм. The output voltage of the first stage is defined as $U_{out} = R_{TR} i_c = -6 \cdot 10^6 (-100 \cdot 10^{-9}) = +0.6$ V, i.e. the first cascade at signal frequencies converts a current equal to -100 nA into a positive pulse with an amplitude of +0.6 V.

The second stage is also implemented using the inverting OA circuit with voltage feedback. The transfer impedance of this circuit is equal to $R_{TR} = -R_{OA} = -R_7$. The output voltage of the second stage is determined as $U_{out} = R_{TR} \times i_c = -20 \times 300 \times 10^{-6} = -6$ V, i.e. the second stage at the signal frequencies converts a current equal to 300 μA into a negative pulse with an amplitude of -6 V. The amplitude of the output signal from the second stage is determined by the potentiometer $R_5$ and can also vary widely from -0,6 В (than $R_5 = 20$ кОм, when the second cascade operates as an inverter) to a maximum of about – 10 V.

Thus, as a result of amplification at the second amplifier stage output, the pulse amplitude changes from zero to the maximum amplitude (can reach ~10 volts). To prevent self-oscillation of the second stage when capacitive loads are connected to the analog output, a resistor $R_6$ is included in the stage's output circuit. Even with large capacitive loads, a sufficiently large phase margin is created, ensuring stable operation of the amplifier stage.

A general-purpose comparator is used to select analog signals above a certain threshold and to convert the input information into a single binary message. Since the input pulse signal has a steep rise and a rapid fall near the discrimination level, there is no need to apply special measures to prevent output signal chatter. The comparator [Horowitz & Hill, 1980] operates with a constant reference voltage of 1 volt. The output stage of the comparators used is implemented using an open collector circuit. This enables the implementation of current signal transmission for communication with computing devices.

### 4. Data collection and processing system

With the widespread adoption of microprocessor technology, an approach in which each detector is equipped with its own built-in data acquisition system becomes possible. This system has few interconnections, which leads to increased reliability and autonomy of individual detectors. Detector autonomy is important because marine expeditionary measurements require multiple detectors located in different laboratories on a vessel, sometimes widely separated.

The requirements for the data acquisition system are fairly simple. Secondary cosmic radiation in the deep atmosphere is essentially background radiation (<100 pps sec-1). The data acquisition system operates in simple pulse counting mode. The primary requirement for the data acquisition system is reliability, as the equipment must operate continuously in limited access mode (>180 days).

The recording system is based on an Arduino-Uno microcontroller [Mikhalko et al. 2021]. Output pulses of ~10 μs duration, generated by an open-collector amplifier-discriminator, are fed to the microcontroller's '0' or '1' interrupts, configured to trigger on the rising edge.

In addition to the detector count rate, a number of other parameters are recorded with the same time resolution. A BMP280 sensor is used to record atmospheric pressure and internal ambient temperature. (±0.12 *hPa*, ±0.01 °*C*). A waterproof digital sensor DHT22 is used as an external temperature and humidity sensor (its parameters are: (–40 ÷ 125) ±0.5 °*C*; (0 ÷100)±2 %).

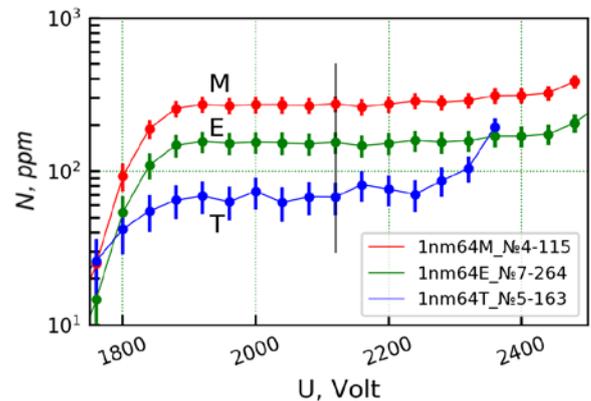

**Figure 3.** Counting characteristics of the detectors: 1NM64M – lead neutron monitor, 1NM64E – epithermal neutron detector, 1NM64T – thermal neutron detector (single counter SNM-15). The vertical line indicates the operating voltage of the counters, 2120 volts.

To obtain geographic coordinates, a GPS/Glonass/Beidou/Galileo receiver module for an Arduino project based on the ATGM336H chip is used. The modules communicate with each other using a UART hardware protocol for serial communication between devices. This allows the GPS antenna to be extended up to 15 meters, which is important in marine conditions.

To maintain accurate time, a DS3231 real-time clock (accuracy ±5 seconds per month) is used, periodically synchronizing with world time based on data from the GPS/Glonass/Beidou/Galileo receiver.
Data is recorded to a 32 GB microSD card, which is sufficient to store a year's worth of data at a minute resolution for all 12 data channels. If necessary, data can be transferred to a PC via a USB interface.

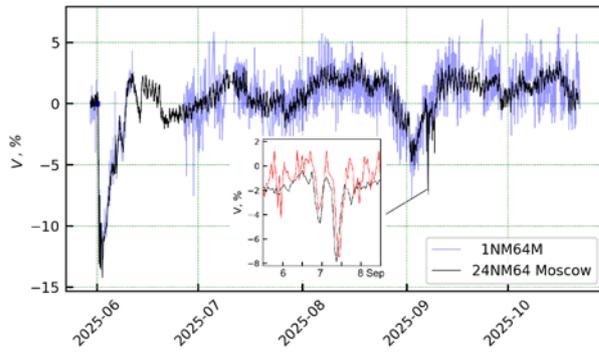

**Figure 4.** Comparison of variations in the neutron component of cosmic rays in Moscow for lead neutron monitors 1NM64M and standard 24NM64M.

## 5. Main characteristics of detectors

The main characteristics of the neutron detectors are summarized in Table 1. The statistical measurement error of the 1NM64M detector is ~0.8%/hour or ~0.15%/day. Figure 3 shows the count rate versus voltage applied to the counter curve. The plateau length and slope of the count curve are given in Table 1 for each detector under consideration. The operating voltage of the counters is ~2120 volts, the plateau length is >300 volts, and the plateau slope does not exceed 0.05%/V [Naumov et al. 2015]. A BNV-31 ("Aspect") high-voltage power supply unit with a temperature coefficient of -30 $ppm/°C$ is used, which, even with a temperature fluctuation of 30°C, leads to a voltage change at the operating point of only a few volts. An alternative is the BNV-07 ("Aspect") with a temperature coefficient of 50 $ppm/°C$ or the NT-4000P ("Magnitura") unit with a temperature coefficient of 25 $ppm/°C$.

The intensity of cosmic rays decreases with increasing of atmospheric density (atmospheric pressure).

Table 1. Main characteristics of neutron detectors.

|  | 1NM64M | 1NM64E | 1NM64T |
|---|---|---|---|
| Average count rate in May 2025, $min^{-1}$ | 290 | 160. | 70 |
| Statistical acccuracy, %/$hour$ | 0.8%/$hour$ | ~1%/$hour$ |  |
| Length of the plateau of the counting curve, $V$ | 500 | 450 | 350 |
| Slope of the plateau of the counting curve, %/$V$ | 0.04 | 0.04 | 0.05 |
| Forbush effect amplitude on June 1, 2025, % | 13 | 12 | 12 |
| Barometric coefficient, β %/ $hPa$ | 0.74±0.02 | 0.76±0.04 |  |
| The effect of air humidity, ε %/$g$ /$m^3$ | 0.034±0.005 | 0.056±0.009 |  |
| The proportion of reference station variations in the observed variations, δ | 0.59±0.07 | 0.57±0.07 |  |

Air humidity also affects the detection of cosmic rays, as atmospheric water vapor absorbs secondary neutron radiation. To correct for changes in cosmic ray intensity, taking into account the barometric and humidity effects of cosmic rays, coefficients are used that determine the percentage change in particle intensity with a one-unit change in pressure and absolute humidity [Kobelev et al. 2024]. Table 1 shows the barometric and humidity coefficients.

## 6. Results of continuous monitoring

Figure 4 compares variations of the cosmic ray neutron component for Moscow from May 2025 to the present for two 1NM64M lead neutron monitors and the standard 24NM64 neutron monitor.

The period under consideration is close to solar maximum, when 19 Forbush decreases were observed, three of which exceeded 8%; details can be found at https://crst.izmiran.ru/w/feid. The tab shows the Forbush decrease on September 7, 2025.

Figure 5 (left panel) shows the results of the large Forbush decrease on June 1, 2025, for three experimental neutron detectors and compares them with data from the standard Moscow neutron monitor 24NM64. Based on the statistical data, the data from the 1NM64M expeditionary neutron monitor and the 1NM64E epithermal neutron detector are in good agreement with the standard detector. Furthermore, Figure 5 clearly shows the epithermal neutron humidity effect, which was clearly evident on June 8 after a light rain (13 UTC, 6 mm of water; marked with a rain symbol in the figure).

Thermal neutron monitoring was conducted only for a few days from June 1-3, but at a very opportune moment—during the onset and evolution of the Forbush decrease.

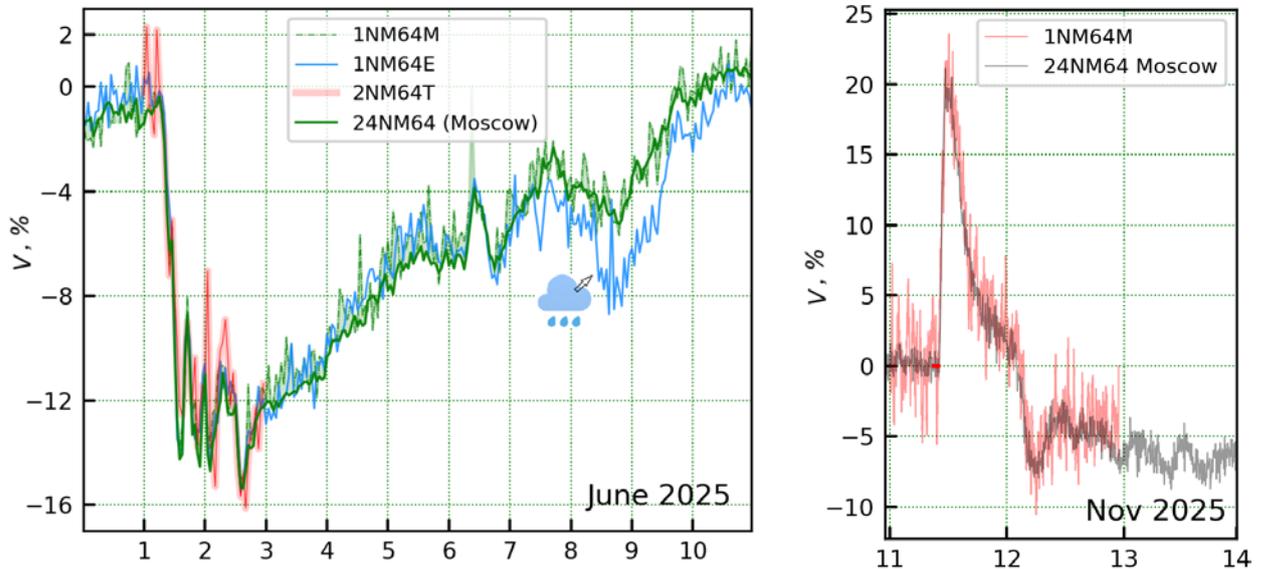

**Figure 5.** Left panel – behavior of neutron component variations for the Forbush decrease on June 1, 2025, recorded by the 1NM64M expeditionary neutron monitor, the 1NM64E epithermal neutron detector, and the 1NM64T thermal neutron detector with data from the Moscow 24NM64 neutron monitor.
Right panel – Ground-based solar cosmic ray enhancement on November 11, 2025 (GLE c077).

It should be noted that, as expected, the Forbush effect amplitudes for all components are similar for the different detectors, since the effect is determined by the modulation of the primary flux in the heliosphere.

Figure 5 (right panel) shows the result of a rare ground-based increase of solar cosmic rays on November 11, 2025 (GLE C077). Taking into account the statistics, good agreement is also observed between the data from the expedition's 1NM64M and the standard neutron monitor.

**7. Conclusion**

The detector proposed in this paper is designed for continuous monitoring of background neutron radiation during expeditionary measurements under weight and size constraints. The detector weighs 2000 kg and has an area of 1 m². An alternative method for forming a lead moderator using lead blocks is presented, and their optimal dimensions are determined.

The developed system comprises the necessary modules for conducting the experiment. It includes a reformatted neutron monitor, an electronics unit, low- and high-voltage power supplies, and an integrated data acquisition, processing, and storage system. The characteristics of the 1NM64M neutron detector were experimentally determined. Detectors can be easily cascaded, if necessary, improving statistical accuracy.

To obtain coordinates, a GPS/GLONASS/Beidou/Galileo receiver module for an Arduino project was used, which is connected to the data collection system via a UART hardware communication protocol. This allows the GPS antenna to be positioned at a distance of up to 15 meters, which is important for communication conditions. Weather data (atmospheric pressure, surface temperature, and air slope) are recorded along the vessel's route. It is worth noting that this set of modules meets the requirements of this experiment and can be easily modified or expanded.

## Acknowledgments

The authors are grateful to the participants of the NMDB project (www.nmdb.eu). This work is being conducted within the framework of the Russian National Ground Network of Cosmic Ray Stations (USI unique scientific installation) (https://ckp-rf.ru/catalog/usu/433536) and the Ministry of Science Marine Scientific Research Project 2025–2028.